\documentclass[
journal=jctcce, 
doi=true,
manuscript=article,layout=onecolumn
]{achemso}

\usepackage[version=3]{mhchem} 

\usepackage{dcolumn}
\usepackage{bm}
\usepackage{amsfonts}
\usepackage{bbm}
\usepackage{todonotes}
\usepackage{tikz}
\usepackage{cool}
\usepackage{dirtytalk}
\usepackage{physics}
\usepackage[fleqn]{mathtools}
\usepackage{isomath}
\usepackage{lmodern}

\def\vector#1{\vectorsym{#1}}

\def\der#1{\textrm{d} #1~}
\def\avg#1{\left< #1 \right>}

\newcommand{\bianca}{\renewcommand\NAT@open{[}\renewcommand\NAT@close{]}}

\title{The Anisotropy of the Proton \\Momentum Distribution in Water}

\author{Venkat Kapil}
\email{venkat.kapil@epfl.ch}
 \affiliation{Laboratory of Computational Science and Modelling, Institute of Materials, Ecole Polytechnique F\'ed\'erale de Lausanne, Lausanne, Switzerland}
 
\author{Alice Cuzzocrea}
\affiliation{Laboratory of Computational Science and Modelling, Institute of Materials, Ecole Polytechnique F\'ed\'erale de Lausanne, Lausanne, Switzerland}

\author{Michele Ceriotti}
\affiliation{Laboratory of Computational Science and Modelling, Institute of Materials, Ecole Polytechnique F\'ed\'erale de Lausanne, Lausanne, Switzerland}%

\date{\today}

\begin{document}
\begin{tocentry}
\centering\includegraphics[width=\textwidth]{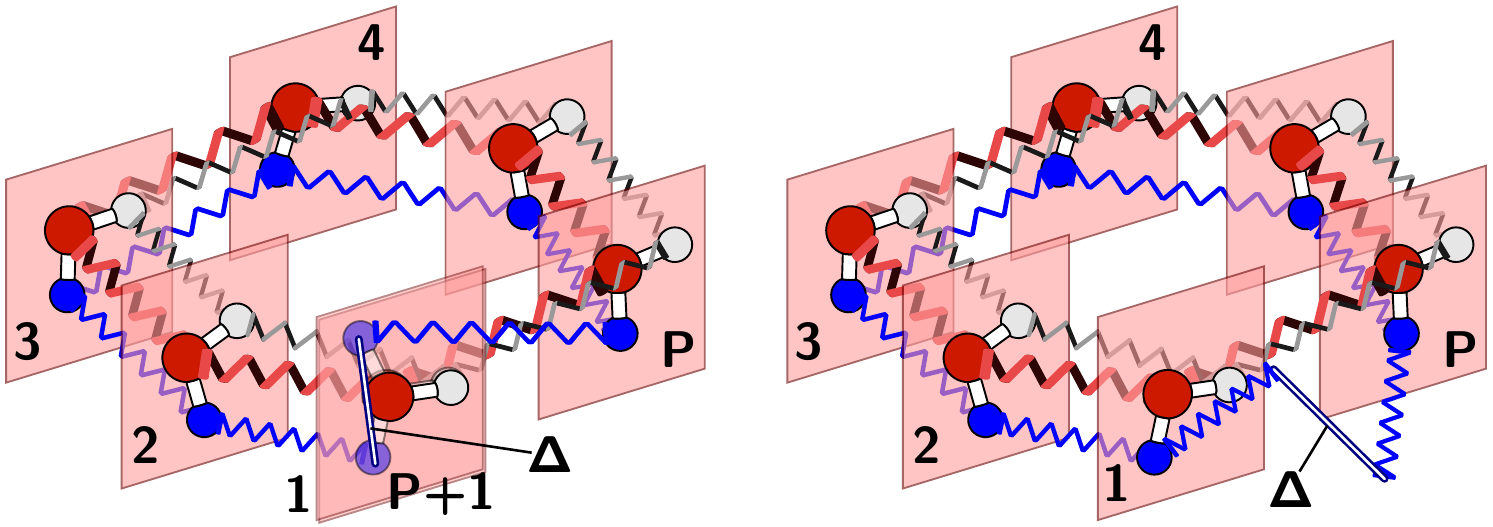}
\end{tocentry}

\begin{abstract}

One of the many peculiar properties of water is the pronounced deviation of the proton momentum distribution from Maxwell-Boltzmann behaviour. 
This deviation from the classical limit is a manifestation of the quantum mechanical nature of protons. Its extent, which can be probed directly by Deep Inelastic Neutron Scattering (DINS) experiments, gives important insight on the potential of mean force felt by H atoms. 
The determination of the full distribution of particle momenta, however, is a real \emph{tour de force} for both experiments and theory, which has led to unresolved discrepancies between the two. 
In this Letter we present comprehensive, fully-converged momentum distributions for water at several thermodynamic state points, focusing on the components that cannot be described in terms of a scalar contribution to the quantum kinetic energy, and providing a benchmark that can serve as a reference for future simulations and experiments. 
In doing so, we also introduce a number of technical developments that simplify and accelerate greatly the calculation of momentum distributions by means of atomistic simulations.
\end{abstract}

\maketitle

\section{Introduction}

Aqueous systems exhibit many distinctive properties\cite{Gallo1543}, many of which are affected by the quantum nature of protons, even at room temperature and above\cite{ceri+16cr}.One of the most evident signatures of the quantum fluctuations of nuclei is the deviation of the momentum distribution of protons $n(\mathbf{p})$ from the classical Maxwell-Boltzmann limit. 
This is essentially a consequence of the fact that position and momentum do not commute, making the distribution of momentum  depend on the local potential felt by the proton. This quantity can be measured directly through Deep Inelastic Neutron Scattering (DINS) experiments\cite{reit+02prb,reit+04bjp,andr+05advp,andr+17ap,Andreani2017}. 
However, the interpretation of these experiments is not straightforward because the dependence of the momentum on the potential is not a trivial one, and because the spherically-averaged momentum distribution contains relatively little information, and so high quantitative accuracy of measurements and simulations is necessary to reach a compelling comparison.  
Early reports of large anomalies in the temperature dependence of proton kinetic energy in supercooled water~\cite{piet+08prl,flam+09jcp} have not been reproduced in path integral simulations using empirical forcefields~\cite{rami-herr11prb}, and have been considerably reassessed in subsequent measurements~\cite{andr+16jpcl}.
An early study of this discrepancy based on vibrational self-consistent field calculations came to the conclusion that softening of the OH stretch due to electrostatic interactions could not reproduce the experimental momentum distribution at room temperature and below~\cite{burn+11jcp}. The model, however, was giving a generally poor description of quantum fluctuations, and even failed to account for the sign of the change in kinetic energy upon condensation. 
More recently, a concerted effort between experiment~\cite{andr+16jpcl} and modelling~\cite{chen+16jpcl} has shown that widely different interatomic water potentials reproduce to a very good accuracy the equilibrium isotope fractionation (that is directly related to the quantum kinetic energy~\cite{chen+16jpcl}), and that their proton kinetic energy differs less than 2meV/atom among each other.
Experiments close to the melting point agree very well with these estimates. Disagreement still exists, however, for room-temperature water, for which experiments consistently measure a kinetic energy which is about 7\% lower than predicted by modelling. 
Furthermore, the quantum kinetic energy is not sufficient to characterize fully the quantum momentum distribution, that results from the combination of several vibrational modes with different zero-point energies.

A more complete comparison requires the measurement and the evaluation of the full, anisotropic particle momentum statistics.
The standard methods for the computation of $n(\mathbf{p})$ are  \say{open}\cite{morr+07jcp,pant+08prl} and \say{displaced}\cite{lin+10prl} Path Integral Molecular Dynamics (PIMD) simulations which include exactly the quantum effects on the motion of nuclei on an \textit{ab initio} potential energy surface. 
However, these simulations have a particularly high computational cost. In the case of open PIMD, the estimator used to calculate the momentum distribution allows to accumulate statistics for a single particle out of the entire simulation, and as a consequence requires very long trajectories to obtain converge averages.
Displaced PIMD computes the distribution by estimators based on free energy perturbation  -- that can be affected by large statistical artifacts~\cite{ceri+12prsa} --  or on thermodynamic integration -- that require multiple trajectories and allows to sample one single particle. 
A few \emph{ab initio} simulations exist that show qualitative agreement with experiments for the full momentum distribution~\cite{morr-car08prl,ceri+10prb,flam+12jcp}. However, as discussed above, clear discrepancies still exist\cite{pant+08prl,andr+16jpcl,flam+12jcp}. The integrated nature of the signal, and the subtle dependence on thermodynamic conditions, call for a quantitative benchmark to be established.
\begin{figure}[t]
\centering\includegraphics[width=\textwidth]{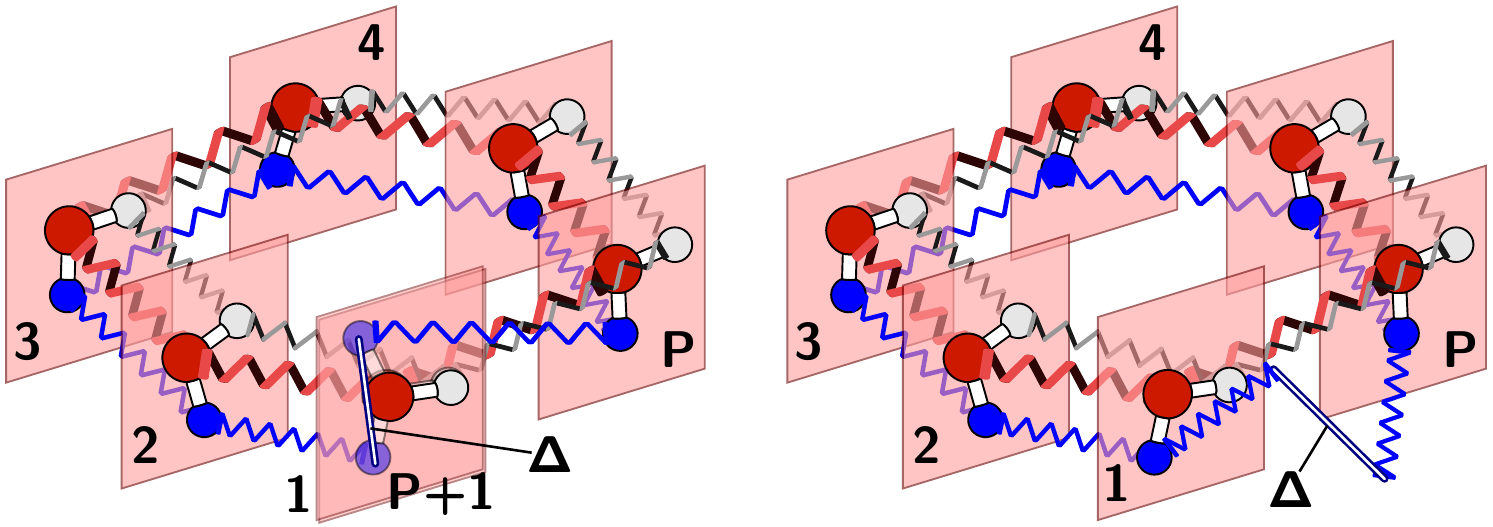}
\caption{The discretization of the imaginary time path integral using a uniform grid starting from $\tau=0$ to  $\tau=\beta \hbar$ (left) gives the standard open PIMD Hamiltonian while the alternate discretization (right) makes it possible to integrate out the ends, yielding a Hamiltonian which describes an open paths with $P$ replicas.
\label{fig:alt_split}}
\end{figure}
To achieve this, we introduce an improved scheme to extract the full particle momentum distribution from PIMD, and use it to present a comparison between three different water models and the most recent DINS measurements.

\section{Methods}
We take as starting point a well-established expression for the canonical particle momentum distribution of the $i$-th atom in a system composed of $N$ atoms at inverse temperature $\beta$, in terms of the Fourier transform of the off-diagonal components of the one-particle density matrix:
\begin{align}
n\left(\vector{p}\right) = \left(2 \pi \hbar \right)^{-3} \int \der{\vector{\Delta}} e^{-i \hbar^{-1}\vector{p}\cdot \vector{\Delta}} ~\rho_i(\vector{0},\vector{\Delta}),
\end{align}
with
\begin{equation}
\rho_i(\vector{0},\vector{\Delta}) =  Z^{-1} \int \der{\{\vector{q}\}} \mel{\vector{q}_{1}, \dots \vector{q}_{i}, \dots, \vector{q}_{N}}{e^{-\beta \hat{H}}}{\vector{q}_{1}, \dots, \vector{q}_{i}-\vector{\Delta}, \dots, \vector{q}_{N}}. \label{eq:n-delta}
\end{equation}
Eq.~\eqref{eq:n-delta} can be computed performing a second order Trotter splitting of the Boltzmann operator into $P$ factors,
$    e^{-\beta \hat{H}} \approx  \left[e^{-\beta_P \frac{\hat{V}}{2}} ~e^{-\beta_P \hat{T}} ~e^{-\beta_P \frac{\hat{V}}{2}}\right]^{P}  $
and inserting complete sets of position states between each pair of factors. 
This yields the familiar isomorphism with a classical ring-polymer partition function~\cite{feyn-hibb65book,chan-woly81jcp,parr-rahm84jcp} sampled at $\beta_P = \beta/P$, and the equivalence between the single-particle density matrix $\rho_i(\vector{0},\vector{\Delta})$ and the end-to-end distribution for the $i$-th path $N(\vector{\Delta})$.
Since the end points of all but one ring polymer coincide, one has to introduce an additional replica just to evaluate the potential for two configurations that differ only by the positions of the $i$-th particle, and that are each weighted by a factor $1/2$ (Figure~\ref{fig:alt_split}). 
Besides the slight computational overhead, the presence of the extra replica for the target species and the scaling of the potential makes the implementation of this Hamiltonian in an existing PIMD code somewhat cumbersome. 
To avoid this inconvenience, we use in our implementation in the i-PI code~\cite{ceri+14cpc} an alternative second order splitting~
$e^{-\beta \hat{H}} \approx \left[e^{-\beta_P \frac{\hat{T}}{2}} ~e^{-\beta_P \frac{\hat{V}}{2}}\right] \left[e^{-\beta_P \frac{\hat{V}}{2}} ~e^{-\beta_P \hat{T}} ~e^{-\beta_P \frac{\hat{V}}{2}}\right]^{(P-1)} \left[e^{-\beta_P \frac{\hat{V}}{2}} ~e^{-\beta_P \frac{\hat{T}}{2}}\right]$. This makes it possible to compute the end-to-end distribution $N(\vector{\Delta})$ by sampling the Hamiltonian of a truncated ring polymer
\begin{align}
H_{P} = \sum_{k=1}^{P} \left[\sum_{j=1}^{N} \frac{1}{2} m_j^{-1}[\vector{p}_{j}^{(k)}]^2 + \sum_{j=1}^{N}\frac{1}{2} m_j \omega_P^2 [\vector{q}_{j}^{(k)} - \vector{q}_{j}^{(k+1)}]^2+ V(\vector{q}^{(k)}) \right] -  \frac{1}{2} m_i \omega_P^2 [\vector{q}_{i}^{(1)} - \vector{q}_{i}^{(P)}]^2 
\end{align}
in which all the replicas experience the full physical potential and both open and closed replicas are represented by $P$ beads. 
Implementing Molecular Dynamics or Monte Carlo sampling for this Hamiltonian in an existing closed-path code is trivial since it just requires the modification of the normal mode (or staging) transformation for the target species~\cite{ceri+10jcp}.

The attentive reader will have noticed that the alternative Trotter splitting contains two additional free-particle propagators, that describe the fluctuations of the end-points of the path around the first and the $P$-th bead. 
These fluctuations can be evaluated analytically, and correspond to a Gaussian convolution of the 1-st-to-$P$-th bead distribution. 
In practice, in order to estimate  $N(\vector{\Delta})$ one simply needs to compute the histogram of $\vector\Delta$ using a kernel function that depends parametrically on $\beta$, $m$ and $P$:
\begin{align}
N(\vector{\Delta})& = \lim_{P \to \infty}\avg{G^{\text{3D}}(\vector{\Delta}, \vector{q}_{i}^{(1)}-\vector{q}_{i}^{(P)})}_{{H}_P}, \label{eq:n-g-3d}
\end{align}
with 
\begin{equation}
G^\text{3D}(\vector{x},\vector{x}') = \left(\sqrt{2 \pi \sigma_P}\right)^{-3} e^{-\frac{(\vector{x}- \vector{x'})^2}{2\sigma_P^2}} \quad \text{and} \quad \sigma_P = \sqrt{m_i^{-1} \beta_P \hbar^2}
\end{equation}
When computing $n(\vector{p})$ in an isotropic system such as liquid water, one is generally interested in the spherical average of the end-to-end distribution, $N(\Delta)$, because the spherically-averaged momentum distribution can be obtained from it as 
\begin{equation}
n(p)= \int \der{\Delta} 4 \pi \Delta^2 N(\Delta) \frac{\Sin{p \Delta}}{p \Delta}. \label{eq:n-g-r}
\end{equation}
$N(\Delta)$ can also be obtained by computing a histogram with an appropriate kernel, 
\begin{equation}
N(\Delta) = \lim_{P \to \infty}\left<G^{\text{r}}(\Delta,\left|\vector{q}_i^{(1)}-\vector{q}_i^{(P)}\right|)\right>_{H_P}. \label{eq:n-d-rad}
\end{equation}
The expression for $G^{\text{r}}(x,x')$ is given in the SI, where we also discuss the construction of a virial-like scaled gradient (SG) estimator for the derivative of the end-to-end distribution: 
\begin{align}
& \frac{\der{N(\Delta)}}{\der{\Delta}} = -\lim_{P \to \infty}\avg{\vector{g} \cdot \left[\frac{\vector{q}_i^{(1)}-\vector{q}_i^{(P)}}{\abs{\vector{q}_i^{(1)}-\vector{q}_i^{(P)}}}\right] ~G^{\text{dr}}\left(\Delta, \abs{\vector{q}_i^{(1)}-\vector{q}_i^{(P)}}\right)}_{{H}_P}
\label{eq:sg-rad}
\end{align}
where $\vector{g} = \left[\frac{\sigma_P^{-2}}{1-P^{-1}} \left[\vector{q}_i^{(1)}-\vector{q}_i^{(P)}\right] - \sum_{k=1}^{P} \beta_P \lambda_k \frac{\partial V(\vector{q}^{(k)})}{\partial \vector{q}_i^{(k)}} \right]$, where $\lambda_k$ is an arithmetic progression, beginning and ending at $-\frac{1}{2}$ and $\frac{1}{2}$. The SG estimator bears some formal similarities with the displaced-path estimators of Lin et al.~\cite{lin+10prl} but can be used in an open path calculation to obtain an independent estimate of the end-to-end distribution at no additional cost. In our simulations, this route consistently showed much smaller statistical errors than the direct evaluation of~\eqref{eq:n-d-rad}.
\begin{figure}[t]
\centering\includegraphics[width=0.5\columnwidth]{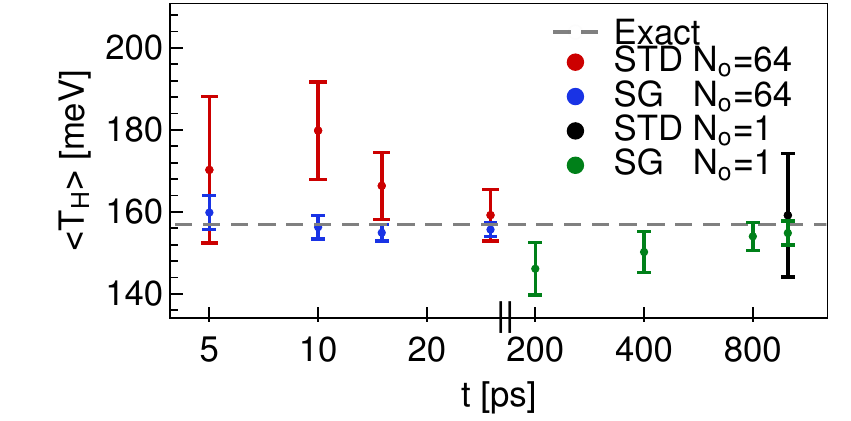}
\caption{ The convergence of the kinetic energy of a proton in supercooled q-TIP4P/f water at 271K from open PIMD simulations using 64 beads calculated from the momentum distribution using the scaled gradient (SG) estimator (green). As a reference for comparison the value calculated from the standard estimator (black) with 1ns of trajectory and the exact result (dashed) from closed PIMD are also plotted. The red and the blue points are obtained using the standard and the scaled gradient estimators from simulations where one proton per molecule is represented by an open path. 
\label{fig:eff_2}}
\end{figure}

To test the convergence of the regular and the scaled gradient estimators we ran simulations of of 64 molecules of q-TIP4P/f water at 271K using $64$ imaginary time slices. The distribution of momentum of a proton in water can essentially be described as an anisotropic Gaussian~\cite{andr+05advp,ceri-mano12prl,flam+12jcp}, so we can use as a convergence check the comparison between the quantum kinetic energy computed from a closed-path simulation and that obtained from the second moment of $n(p)$. 
As shown in figure \ref{fig:eff_2}, the scaled gradient estimator (which can be obtained from an open path simulation at virtually no computational overhead) shows an error which is approximately 5 times smaller than that associated with the standard estimator. 
For simulations with a single open path, 1ns-long simulations are needed to approach a statistical error sufficient to discriminate between different water molecules, whereas more than 20 would be needed without using the derivative estimator. 
Fig.~\ref{fig:eff_2} also demonstrates that the approximation introduced in Ref.~\cite{morr+07jcp} -- that is, opening one path per water molecule -- leads to negligible systematic error, and once combined with the derivative estimator allows reaching a statistical error of about 2 meV with trajectories that are only 10 ps long. 
\begin{center}
\begin{table*}
 \begin{tabular}{l | c c c c }
 \hline\hline
 Phase & B3LYP+D3 & MBPOL & q-TIP4P/f & Exp\\
  & [meV] & [meV] & [meV] & [meV]\\ [0.5ex]
 \hline
 271 K [SW] & 154.4 & 157.2 & 156.9 & $[156 \pm 2]$\cite{andr+16jpcl}\\
 271 K [$\text{I}_\text{h}$] & 154.5 & 158.1 & 157.4 & $[157 \pm 2]$\cite{andr+16jpcl} $[156 \pm 2]$\cite{flam+12jcp} \emph{[158 $\pm$ 4] \cite{sene+13jcp}} \\
 296 K [W] & 155.2 & 158.5 & 157.6 & $[146 \pm 3]$\cite{pant+08prl} \emph{[150 $\pm$ 4] \cite{sene+13jcp}} \\
 300 K [W] &155.4 &158.6 &  157.9& $[146 \pm 3]$\cite{andr+16jpcl}\\
 \hline
 \hline
\end{tabular}
 \caption{Mean proton kinetic energy of various phases of B3LYP+D3, MBPOL and q-TIP4P/f water from closed PIMD simulations with 64 imaginary time slices. Statistical error bars are witin 0.1 meV. Experimental results are from DINS measurements, except for the INS estimates (in italics) from Ref.~\citenum{sene+13jcp}. \label{table:KE}}
\end{table*}
\end{center}

\section{Results}

Having demonstrated the potential of the scaled gradient estimator, we turn our attention to the case of the momentum distribution of various phases of water at room temperature and below. 
Experiments performed in this regime have observed a non monotonic dependence of the kinetic energy across the temperature of maximum density of water\cite{flam+09jcp}. They indicate that
supercooled water possesses an excess proton kinetic energy\cite{andr+16jpcl,sene+13jcp}, about $7$\%{}  higher than that of room temperature water.
In order to investigate the source of this anomaly, we use three very successful water models- q-TIP4P/f\cite{habe+09jcp}, NN-B3LYP+D3\cite{kapi+16jcp2} and MBPOL\cite{medders+jctc2013} - to estimate the proton momentum distribution and the mean proton kinetic energy respectively using open and closed PIMD simulation for ice and supercooled water at $271$~K and liquid water at $296$~K and $300$~K at their experimental densities. 
Details about the simulations are given in the SI.
To avoid the huge computational cost of \emph{ab initio} path integral simulations, we use a neural network~\cite{behl-parr07prl,mora+16pnas} fit to reference density functional theory calculations using the B3LYP hybrid functional~\cite{Stephens1994} and D3 empirical Van der Waals corrections~\cite{grimme-jcp10-dftd_disp_functional}, that was shown in previous publications to provide very accurate estimates of the quantum properties of the first-principles reference at a fraction of the cost~\cite{kapi+16jcp2,chen+16jpcl}. 
Since the MBPOL potential also has a cost that is not negligible, we fit a short-range NN potential\cite{behl-parr07prl,imba+18jcp} to MBPOL energy and forces (parameters of the fit provided in the SI) 
and use it together with the i-PI implementation~\cite{kapi+16jcp} of multiple time step molecular dynamics~\cite{tuck+92jcp} and ring polymer contraction~\cite{mark-mano08cpl}, performing a full MB-POL evaluation every 4 steps, and on a contracted ring polymer with 8 beads , which are very conservative values that still allows us to achieve a 20$\times$ speedup in terms of wall-clock time.

\begin{figure*}[t]
\centering\includegraphics[width=\textwidth]{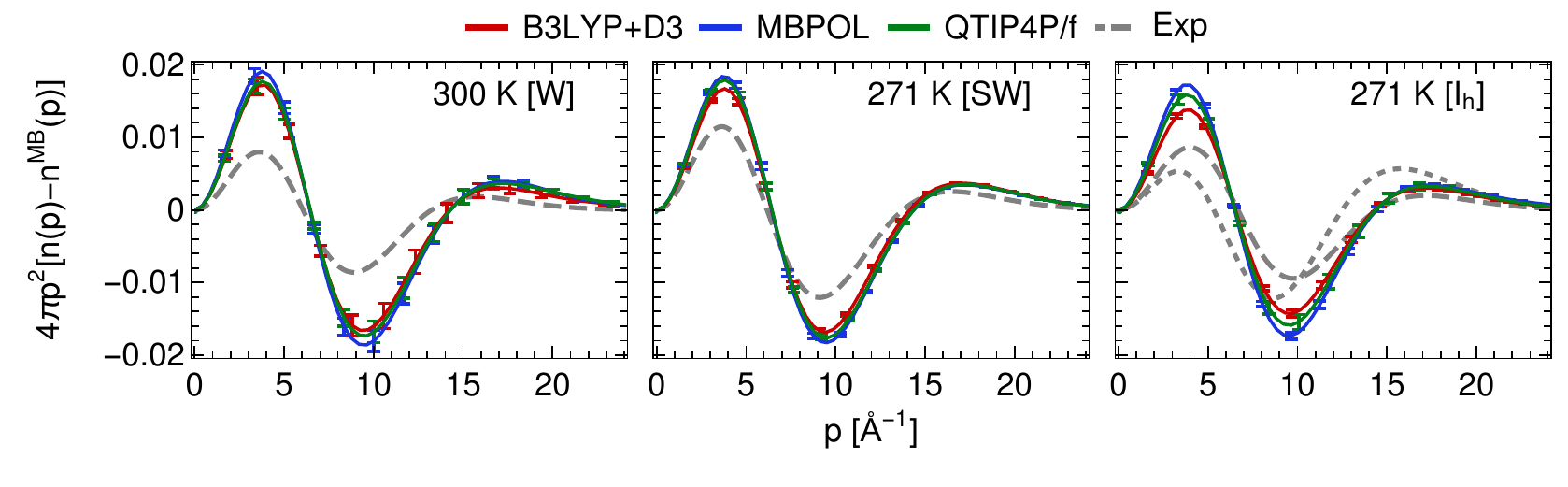}
\caption{ The difference between the radial proton momentum distribution and the Maxwell-Boltzmann distribution with the same second moment for  B3LYP+D3 (red), MBPOL (blue),  q-TIP4P/f (green) and experimental water (grey). The dashed lines correspond to a recent DINS experiment\cite{andr+16jpcl} on room temperature, supercooled and hexagonal ice. The dotted line corresponds to a distribution constructed from a Inelastic Neutron Scattering (INS) experiment for hexagonal ice at 271 K\cite{sene+13cp}. 
\label{fig:np}}
\end{figure*}

We first consider the mean proton kinetic energies listed in Table \ref{table:KE}. Our results for supercooled water and ice are in excellent agreement with the experiment\cite{andr+16jpcl} and the values calculated at the triple point in an earlier work\cite{chen+16jpcl}. 
Contrary to the experimental observation, however, the increase in kinetic energy between the supercooled  liquid and room-temperature conditions suggests that the proton doesn't experience any softening of free energy along the proton transfer coordinate, and that the change in kinetic energy is consistent with a simple increase in thermal excitations of the low-frequency degrees of freedom.

\begin{table*}
 \begin{tabular}{l | c c c c }
 \hline\hline
 Phase & B3LYP+D3 & MBPOL & q-TIP4P/f & Exp\\
  & [$\AA^{-1}$] & [$\AA^{-1}$] & [$\AA^{-1}$] & [$\AA^{-1}$]\\ [0.5ex]
 \hline
 271 K [SW] & (3.1,  3.9,  7.0) & (3.0, 4.0,  7.1) & (3.0,  4.0,  7.0) & (3.7, 4.3, 6.6)\\
 271 K [$\text{I}_\text{h}$] & (3.4, 3.9,  6.9) & (3.1,  3.9,  7.1) & (3.1,  4.0,  7.0) & (2.9, 5.0, 6.5) \\
 300 K [W] &(3.0,  4.1,  6.9)& (3.0 , 3.9,  7.2)&(3.0  4.0  7.1)& (3.1, 5.3, 5.8)\\
 \hline
 \hline
\end{tabular}
\caption{Mean components of the proton kinetic energy along the $x,y,z$ directions for various phases of B3LYP+D3, MBPOL and q-TIP4P/f water, obtained by fitting the spherical average of an anisotropic Gaussian  to the momentum distribution calculated from open PIMD simulations with 64 imaginary time slices. The $x$ direction is along the hydrogen bond and $y,z$ lie on the plane perpendicular to it. The residuals of the fit are of the order of $10^{-3}$. Experimental results for ice and supercooled water are from a recent DINS measurement\cite{andr+16jpcl} while those for room temperature water were obtained from a private communication\cite{pcomm01}. \label{table:KE3D}}
\end{table*}

In order to provide a more complete picture that can help reconcile simulations and experiments, we then proceed to evaluate and compare the full proton momentum distribution. 
Given that the shape of $n(p)$ contains an overwhelming dependence from the mean kinetic energy, we decided to show in Fig.~\ref{fig:np} the difference between the momentum distribution and the Maxwell-Boltzmann distribution corresponding to the quantum kinetic energy reported in table \ref{table:KE}. 
This particular representation highlights the anisotropy introduced by the zero point energy of the collective modes. We find that distributions calculated by the three models are very close to each other. 
This is in line with the results of an earlier work\cite{cheng+jcpl2016} which showed that the kinetic energy is distributed along the stretch, bend and hindered rotation in the ratio $\sim 4:1.5:1$ for all the models, and is consistent with the main vibrational frequencies of water and with inelastic neutron scattering measurements~\cite{sene+13jcp}.
DINS experiments~\cite{andr+16jpcl} are also in good qualitative agreement with the models, although they consistently show a less-pronounced degree of anisotropy.
There is also a small spread in the curves measured at different temperatures and between DINS experiments -- that obtains directly $n(p)$ -- and inelastic neutron scattering data~\cite{sene+13jcp}, that also give qualitative agreement with the computed anisotropy even though it obtains $n(p)$ from an indirect analysis. 
It should be stressed that the nature of the spherical averaging is such that the anisotropic momentum distribution depends rather weakly on the anisotropy of $n(p)$, leading to large error bars on the anisotropy coefficients, and to difficulties in identifying inhomogeneities in the sample~\cite{ceri14jpcs}. 
The computed $n(p)$ can be fitted to an anisotropic Gaussian lineshape~\cite{ceri-mano12prl,flam+12jcp} with essentially no error, which strongly suggests that (at least for room-temperature water) the spherically-averaged distribution does not contain enough information to infer the presence of deviations from a quasi-harmonic description. 
Table~\ref{table:KE3D} shows that the three models are in excellent, quantitative agreement with each other, and small changes as a function of temperature, whereas experimental values are somewhat more erratic, probably due to the difficulty in separating the anisotropic contributions, that are strongly correlated in the fit.

\begin{figure*}[t]
\centering\includegraphics[width=\textwidth]{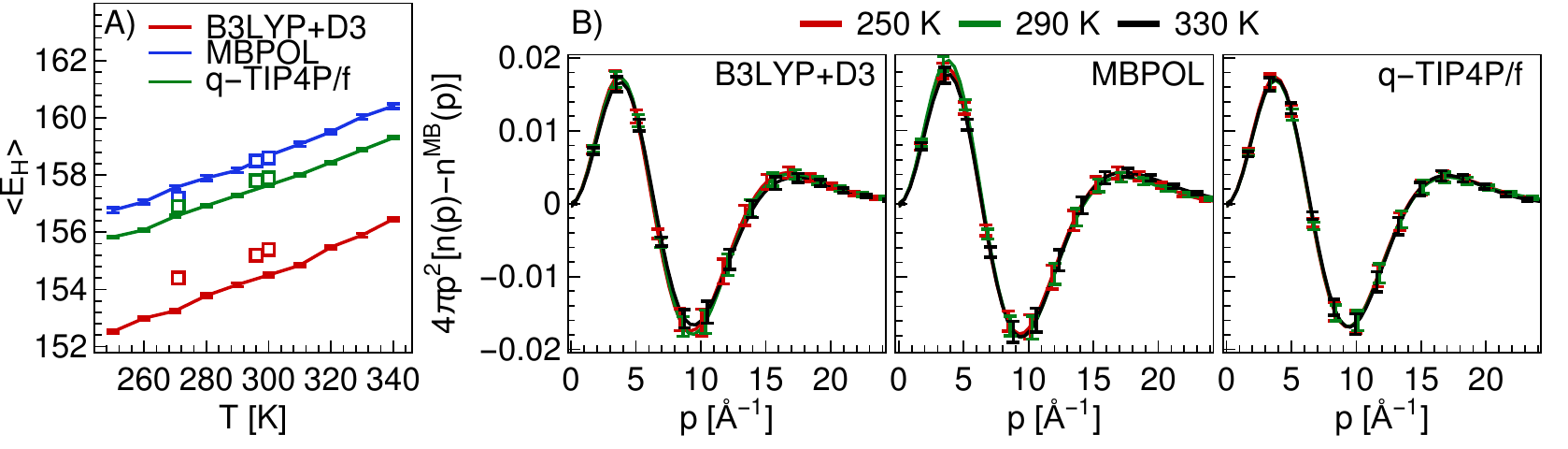}
\caption{Panel A shows the temperature dependence of the proton kinetic energy calculated using closed PIMD simulations that sample the $NPT$ ensemble using B3LYP + D3 (red), MBPOL (blue) and QTIP4P/f (green) water potentials. The square markers correspond to the values obtained from $NVT$ simulations at experimental densities. Panel (B) shows the temperature dependence of the difference between the proton momentum distribution calculated using $NPT$ open PIMD simulations and the Maxwell-Boltzmann distribution with the same mean kinetic energy for B3LYP + D3 (left), MBPOL (middle) and QTIP4P/f (right) water potentials. 
\label{fig:np_TD}}
\end{figure*}

In order to rule out subtle sources of error in the simulations, such as the fact that melting points of different models can vary substantially, and that the equilibrium molar volumes at constant pressure do not always correspond to the experimental density that we used in our NVT simulations, we also performed a scan of the temperature range from $250$~K to $340$~K, running for each water model $10$ additional NpT simulations using both closed paths (to estimate the kinetic energy) and open paths (to estimate the anisotropy). 
After successfully equilibrating the densities, we study the temperature dependence of the kinetic energy and find that all the models display a monotonic increase in the kinetic energy with essentially the same slope (figure \ref{fig:np_TD}) confirming that the disagreement with experiments at room temperature is not an artifact of our simulation protocol. Density effects are also small, as evidenced by the minute differences between NVT and NpT results.
The anisotropic components of $n(p)$ are even more stable across different temperatures, being essentially constant for each model over a temperature range of almost 100K -- which is small compared to the scale of the quantum kinetic energy of protons.

\section{Conclusions}

In summary, we find that the particle momentum distribution of water does not show any trace of anomalous behavior when going from the supercooled limit to temperatures well above room temperature, even when considering the full, anisotropic form of $n(p)$. 
Our analysis is particularly thorough, including three very different water models: a fixed point-charge model, a dissociable machine-learned potential fitted to hybrid density functional theory data, and a state-of-the-art many-body forcefield that has shown consistently quantitative accuracy in reproducing many of the classical and quantum properties of water\cite{medd+15jcp,reddy2017,Paesani2016}.
We repeat our simulations at different thermodynamic conditions, spanning a range of 100K around the experimental melting point, to exclude the possibility that anomalous effects may be shifted to a different temperature due to the inaccuracies of the models. 
The three potentials are in excellent agreement with each other, despite being based on widely different models and physical approximations. The $\approx 2$~meV accuracy of DINS measurements is not sufficient to discriminate between them. 
This should elicit new efforts to improve the accuracy and reliability of DINS experiments, that despite the technical challenges offer a rather unique approach to probe quantum nuclear fluctuations in aqueous environments, complementing other kinds of neutron spectroscopies~\cite{sene+13jcp}. 
We suggest that ice Ih constitutes a very promising system to be used for benchmarking. Measurements on single crystals should give a better handle of the anisotropy of the distribution, and allow for a more stringent benchmark of both atomic-scale models and of measurements of the integrated, radial distribution, that is necessary before a comparison in more challenging conditions~\cite{Gallo1543} can be meaningfully attempted.
Meanwhile, the scaled-gradient estimator we introduce here, together with the efficient and comprehensive implementation in an open-source code that can be interfaced with many electronic-structure packages, and the use of short-range machine-learning potentials that reduce the cost of performing high-end ab initio path integral calculations, should make it possible to further test the accuracy of the atomistic models for this difficult problem.
Efforts in this direction should focus on the remaining disagreement between experiments and models at room temperature. 
As we have shown, the computed particle momentum distribution is essentially the same for increasingly accurate potentials, that also exhibit a kinetic energy that is consistent with equilibrium fractionation data of liquid and gaseous water.
If this last discrepancy cannot be traced to an experimental problem, it might be the signal for an extraordinarily elusive anomalous behavior of this deceptively simple substance.

\section*{Supporting Information}
The derivation of the standard and scaled gradient path integral estimator for the particle momentum distribution using an alternate splitting of the Boltzmann operator, molecular dynamics algorithms for the ring polymer Hamiltonian, the details of the simulations that were performed, benchmarks of the accuracy of the neural network potential, fitting parameters obtained by fitting the particle momentum distribution to a truncated Gauss-Laguerre expansion and a snapshot of the i-PI code with a set of inputs for open path PIMD simulations with the potentials used in the manuscript.

\begin{acknowledgement}
M.C. and V.K. acknowledge financial support by the Swiss National Science Foundation (project ID 200021-159896), and computational time from CSCS under the project IDs s466, s553, and s618. 
A.C. was supported by a MARVEL INSPIRE Potentials Master's Fellowship. MARVEL is a National Center of Competence in Research funded by the Swiss National Science Foundation.
This work was also supported by EPFL through the use of the facilities of its Scientific IT and Application Support Center.
We would like to thank C. Andreani, R. Senesi and G. Romanelli for insightful discussion and for sharing unpublished experimental results.
\end{acknowledgement}

\bibliography{biblio}

\providecommand{\latin}[1]{#1}
\makeatletter
\providecommand{\doi}
  {\begingroup\let\do\@makeother\dospecials
  \catcode`\{=1 \catcode`\}=2 \doi@aux}
\providecommand{\doi@aux}[1]{\endgroup\texttt{#1}}
\makeatother
\providecommand*\mcitethebibliography{\thebibliography}
\csname @ifundefined\endcsname{endmcitethebibliography}
  {\let\endmcitethebibliography\endthebibliography}{}
\begin{mcitethebibliography}{46}
\providecommand*\natexlab[1]{#1}
\providecommand*\mciteSetBstSublistMode[1]{}
\providecommand*\mciteSetBstMaxWidthForm[2]{}
\providecommand*\mciteBstWouldAddEndPuncttrue
  {\def\EndOfBibitem{\unskip.}}
\providecommand*\mciteBstWouldAddEndPunctfalse
  {\let\EndOfBibitem\relax}
\providecommand*\mciteSetBstMidEndSepPunct[3]{}
\providecommand*\mciteSetBstSublistLabelBeginEnd[3]{}
\providecommand*\EndOfBibitem{}
\mciteSetBstSublistMode{f}
\mciteSetBstMaxWidthForm{subitem}{(\alph{mcitesubitemcount})}
\mciteSetBstSublistLabelBeginEnd
  {\mcitemaxwidthsubitemform\space}
  {\relax}
  {\relax}

\bibitem[Gallo and Stanley(2017)Gallo, and Stanley]{Gallo1543}
Gallo,~P.; Stanley,~H.~E. Supercooled water reveals its secrets. \emph{Science}
  \textbf{2017}, \emph{358}, 1543--1544\relax
\mciteBstWouldAddEndPuncttrue
\mciteSetBstMidEndSepPunct{\mcitedefaultmidpunct}
{\mcitedefaultendpunct}{\mcitedefaultseppunct}\relax
\EndOfBibitem
\bibitem[Ceriotti \latin{et~al.}(2016)Ceriotti, Fang, Kusalik, McKenzie,
  Michaelides, Morales, and Markland]{ceri+16cr}
Ceriotti,~M.; Fang,~W.; Kusalik,~P.~G.; McKenzie,~R.~H.; Michaelides,~A.;
  Morales,~M.~A.; Markland,~T.~E. {Nuclear Quantum Effects in Water and Aqueous
  Systems: Experiment, Theory, and Current Challenges}. \emph{Chem. Rev.}
  \textbf{2016}, \emph{116}, 7529--7550\relax
\mciteBstWouldAddEndPuncttrue
\mciteSetBstMidEndSepPunct{\mcitedefaultmidpunct}
{\mcitedefaultendpunct}{\mcitedefaultseppunct}\relax
\EndOfBibitem
\bibitem[Reiter \latin{et~al.}(2002)Reiter, Mayers, and Noreland]{reit+02prb}
Reiter,~G.~F.; Mayers,~J.; Noreland,~J. {Momentum-distribution spectroscopy
  using deep inelastic neutron scattering}. \emph{Phys. Rev. B} \textbf{2002},
  \emph{65}, 104305\relax
\mciteBstWouldAddEndPuncttrue
\mciteSetBstMidEndSepPunct{\mcitedefaultmidpunct}
{\mcitedefaultendpunct}{\mcitedefaultseppunct}\relax
\EndOfBibitem
\bibitem[Reiter \latin{et~al.}(2004)Reiter, Li, Mayers, Abdul-Redah, and
  Platzman]{reit+04bjp}
Reiter,~G.; Li,~J.~C.; Mayers,~J.; Abdul-Redah,~T.; Platzman,~P. {The proton
  momentum distribution in water and ice}. \emph{Braz. J. Phys.} \textbf{2004},
  \emph{34}, 142--147\relax
\mciteBstWouldAddEndPuncttrue
\mciteSetBstMidEndSepPunct{\mcitedefaultmidpunct}
{\mcitedefaultendpunct}{\mcitedefaultseppunct}\relax
\EndOfBibitem
\bibitem[Andreani \latin{et~al.}(2005)Andreani, Colognesi, Mayers, Reiter, and
  Senesi]{andr+05advp}
Andreani,~C.; Colognesi,~D.; Mayers,~J.; Reiter,~G.~F.; Senesi,~R. {Measurement
  of momentum distribution of light atoms and molecules in condensed matter
  systems using inelastic neutron scattering}. \emph{Adv. Phys.} \textbf{2005},
  \emph{54}, 377--469\relax
\mciteBstWouldAddEndPuncttrue
\mciteSetBstMidEndSepPunct{\mcitedefaultmidpunct}
{\mcitedefaultendpunct}{\mcitedefaultseppunct}\relax
\EndOfBibitem
\bibitem[Andreani \latin{et~al.}(2017)Andreani, Krzystyniak, Romanelli, Senesi,
  and Fernandez-Alonso]{andr+17ap}
Andreani,~C.; Krzystyniak,~M.; Romanelli,~G.; Senesi,~R.; Fernandez-Alonso,~F.
  Electron-volt neutron spectroscopy: beyond fundamental systems.
  \emph{Advances in Physics} \textbf{2017}, \emph{66}, 1--73\relax
\mciteBstWouldAddEndPuncttrue
\mciteSetBstMidEndSepPunct{\mcitedefaultmidpunct}
{\mcitedefaultendpunct}{\mcitedefaultseppunct}\relax
\EndOfBibitem
\bibitem[Andreani \latin{et~al.}(2017)Andreani, Senesi, Krzystyniak, Romanelli,
  and Fernandez-Alonso]{Andreani2017}
Andreani,~C.; Senesi,~R.; Krzystyniak,~M.; Romanelli,~G.; Fernandez-Alonso,~F.
  \emph{Neutron Scattering - Applications in Biology, Chemistry, and Materials
  Science}; Elsevier, 2017; pp 403--457\relax
\mciteBstWouldAddEndPuncttrue
\mciteSetBstMidEndSepPunct{\mcitedefaultmidpunct}
{\mcitedefaultendpunct}{\mcitedefaultseppunct}\relax
\EndOfBibitem
\bibitem[Pietropaolo \latin{et~al.}(2008)Pietropaolo, Senesi, Andreani, Botti,
  Ricci, and Bruni]{piet+08prl}
Pietropaolo,~A.; Senesi,~R.; Andreani,~C.; Botti,~A.; Ricci,~M.~a.; Bruni,~F.
  {Excess of Proton Mean Kinetic Energy in Supercooled Water}. \emph{Phys. Rev.
  Lett.} \textbf{2008}, \emph{100}, 2--5\relax
\mciteBstWouldAddEndPuncttrue
\mciteSetBstMidEndSepPunct{\mcitedefaultmidpunct}
{\mcitedefaultendpunct}{\mcitedefaultseppunct}\relax
\EndOfBibitem
\bibitem[Flammini \latin{et~al.}(2009)Flammini, Ricci, and Bruni]{flam+09jcp}
Flammini,~D.; Ricci,~M.~a.; Bruni,~F. {A new water anomaly: the temperature
  dependence of the proton mean kinetic energy.} \emph{J. Chem. Phys.}
  \textbf{2009}, \emph{130}, 236101\relax
\mciteBstWouldAddEndPuncttrue
\mciteSetBstMidEndSepPunct{\mcitedefaultmidpunct}
{\mcitedefaultendpunct}{\mcitedefaultseppunct}\relax
\EndOfBibitem
\bibitem[Ram{\'{i}}rez and Herrero(2011)Ram{\'{i}}rez, and
  Herrero]{rami-herr11prb}
Ram{\'{i}}rez,~R.; Herrero,~C.~P. {Kinetic energy of protons in ice Ih and
  water: A path integral study}. \emph{Phys. Rev. B} \textbf{2011}, \emph{84},
  064130\relax
\mciteBstWouldAddEndPuncttrue
\mciteSetBstMidEndSepPunct{\mcitedefaultmidpunct}
{\mcitedefaultendpunct}{\mcitedefaultseppunct}\relax
\EndOfBibitem
\bibitem[Andreani \latin{et~al.}(2016)Andreani, Romanelli, and
  Senesi]{andr+16jpcl}
Andreani,~C.; Romanelli,~G.; Senesi,~R. {Direct Measurements of Quantum Kinetic
  Energy Tensor in Stable and Metastable Water near the Triple Point: An
  Experimental Benchmark}. \emph{J. Phys. Chem. Letters} \textbf{2016},
  \emph{7}, 2216--2220\relax
\mciteBstWouldAddEndPuncttrue
\mciteSetBstMidEndSepPunct{\mcitedefaultmidpunct}
{\mcitedefaultendpunct}{\mcitedefaultseppunct}\relax
\EndOfBibitem
\bibitem[Burnham \latin{et~al.}(2011)Burnham, Hayashi, Napoleon, Keyes,
  Mukamel, and Reiter]{burn+11jcp}
Burnham,~C.~J.; Hayashi,~T.; Napoleon,~R.~L.; Keyes,~T.; Mukamel,~S.;
  Reiter,~G.~F. The proton momentum distribution in strongly H-bonded phases of
  water: A critical test of electrostatic models. \emph{The Journal of Chemical
  Physics} \textbf{2011}, \emph{135}, 144502\relax
\mciteBstWouldAddEndPuncttrue
\mciteSetBstMidEndSepPunct{\mcitedefaultmidpunct}
{\mcitedefaultendpunct}{\mcitedefaultseppunct}\relax
\EndOfBibitem
\bibitem[Cheng \latin{et~al.}(2016)Cheng, Behler, and Ceriotti]{chen+16jpcl}
Cheng,~B.; Behler,~J.; Ceriotti,~M. {Nuclear Quantum Effects in Water at the
  Triple Point: Using Theory as a Link Between Experiments}. \emph{J. Phys.
  Chem. Letters} \textbf{2016}, \emph{7}, 2210--2215\relax
\mciteBstWouldAddEndPuncttrue
\mciteSetBstMidEndSepPunct{\mcitedefaultmidpunct}
{\mcitedefaultendpunct}{\mcitedefaultseppunct}\relax
\EndOfBibitem
\bibitem[Morrone \latin{et~al.}(2007)Morrone, Srinivasan, Sebastiani, and
  Car]{morr+07jcp}
Morrone,~J.~A.; Srinivasan,~V.; Sebastiani,~D.; Car,~R. {Proton momentum
  distribution in water: an open path integral molecular dynamics study.}
  \emph{J. Chem. Phys.} \textbf{2007}, \emph{126}, 234504\relax
\mciteBstWouldAddEndPuncttrue
\mciteSetBstMidEndSepPunct{\mcitedefaultmidpunct}
{\mcitedefaultendpunct}{\mcitedefaultseppunct}\relax
\EndOfBibitem
\bibitem[Pantalei \latin{et~al.}(2008)Pantalei, Pietropaolo, Senesi, Imberti,
  Andreani, Mayers, Burnham, and Reiter]{pant+08prl}
Pantalei,~C.; Pietropaolo,~A.; Senesi,~R.; Imberti,~S.; Andreani,~C.;
  Mayers,~J.; Burnham,~C.; Reiter,~G. {Proton momentum distribution of liquid
  water from room temperature to the supercritical phase}. \emph{Phys. Rev.
  Lett.} \textbf{2008}, \emph{100}, 177801\relax
\mciteBstWouldAddEndPuncttrue
\mciteSetBstMidEndSepPunct{\mcitedefaultmidpunct}
{\mcitedefaultendpunct}{\mcitedefaultseppunct}\relax
\EndOfBibitem
\bibitem[Lin \latin{et~al.}(2010)Lin, Morrone, Car, and Parrinello]{lin+10prl}
Lin,~L.; Morrone,~J.~A.; Car,~R.; Parrinello,~M. {Displaced Path Integral
  Formulation for the Momentum Distribution of Quantum Particles}. \emph{Phys.
  Rev. Lett.} \textbf{2010}, \emph{105}, 110602\relax
\mciteBstWouldAddEndPuncttrue
\mciteSetBstMidEndSepPunct{\mcitedefaultmidpunct}
{\mcitedefaultendpunct}{\mcitedefaultseppunct}\relax
\EndOfBibitem
\bibitem[Ceriotti \latin{et~al.}(2011)Ceriotti, Brain, Riordan, and
  Manolopoulos]{ceri+12prsa}
Ceriotti,~M.; Brain,~G. a.~R.; Riordan,~O.; Manolopoulos,~D.~E. {The
  inefficiency of re-weighted sampling and the curse of system size in high
  order path integration}. \emph{Proceedings of the Royal Society A:
  Mathematical, Physical and Engineering Sciences} \textbf{2011}, \emph{468},
  2--17\relax
\mciteBstWouldAddEndPuncttrue
\mciteSetBstMidEndSepPunct{\mcitedefaultmidpunct}
{\mcitedefaultendpunct}{\mcitedefaultseppunct}\relax
\EndOfBibitem
\bibitem[Morrone and Car(2008)Morrone, and Car]{morr-car08prl}
Morrone,~J.~A.; Car,~R. {Nuclear Quantum Effects in Water}. \emph{Phys. Rev.
  Lett.} \textbf{2008}, \emph{101}, 17801\relax
\mciteBstWouldAddEndPuncttrue
\mciteSetBstMidEndSepPunct{\mcitedefaultmidpunct}
{\mcitedefaultendpunct}{\mcitedefaultseppunct}\relax
\EndOfBibitem
\bibitem[Ceriotti \latin{et~al.}(2010)Ceriotti, Miceli, Pietropaolo, Colognesi,
  Nale, Catti, Bernasconi, and Parrinello]{ceri+10prb}
Ceriotti,~M.; Miceli,~G.; Pietropaolo,~A.; Colognesi,~D.; Nale,~A.; Catti,~M.;
  Bernasconi,~M.; Parrinello,~M. {Nuclear quantum effects in ab initio
  dynamics: Theory and experiments for lithium imide}. \emph{Phys. Rev. B}
  \textbf{2010}, \emph{82}, 174306\relax
\mciteBstWouldAddEndPuncttrue
\mciteSetBstMidEndSepPunct{\mcitedefaultmidpunct}
{\mcitedefaultendpunct}{\mcitedefaultseppunct}\relax
\EndOfBibitem
\bibitem[Flammini \latin{et~al.}(2012)Flammini, Pietropaolo, Senesi, Andreani,
  McBride, Hodgson, Adams, Lin, and Car]{flam+12jcp}
Flammini,~D.; Pietropaolo,~A.; Senesi,~R.; Andreani,~C.; McBride,~F.;
  Hodgson,~A.; Adams,~M.~a.; Lin,~L.; Car,~R. {Spherical momentum distribution
  of the protons in hexagonal ice from modeling of inelastic neutron scattering
  data.} \emph{J. Chem. Phys.} \textbf{2012}, \emph{136}, 024504\relax
\mciteBstWouldAddEndPuncttrue
\mciteSetBstMidEndSepPunct{\mcitedefaultmidpunct}
{\mcitedefaultendpunct}{\mcitedefaultseppunct}\relax
\EndOfBibitem
\bibitem[Feynman and Hibbs(1964)Feynman, and Hibbs]{feyn-hibb65book}
Feynman,~R.~P.; Hibbs,~A.~R. \emph{{Quantum Mechanics and Path Integrals}};
  McGraw-Hill: New York, 1964\relax
\mciteBstWouldAddEndPuncttrue
\mciteSetBstMidEndSepPunct{\mcitedefaultmidpunct}
{\mcitedefaultendpunct}{\mcitedefaultseppunct}\relax
\EndOfBibitem
\bibitem[Chandler and Wolynes(1981)Chandler, and Wolynes]{chan-woly81jcp}
Chandler,~D.; Wolynes,~P.~G. {Exploiting the isomorphism between quantum theory
  and classical statistical mechanics of polyatomic fluids}. \emph{J. Chem.
  Phys.} \textbf{1981}, \emph{74}, 4078--4095\relax
\mciteBstWouldAddEndPuncttrue
\mciteSetBstMidEndSepPunct{\mcitedefaultmidpunct}
{\mcitedefaultendpunct}{\mcitedefaultseppunct}\relax
\EndOfBibitem
\bibitem[Parrinello and Rahman(1984)Parrinello, and Rahman]{parr-rahm84jcp}
Parrinello,~M.; Rahman,~A. {Study of an F center in molten KCl}. \emph{J. Chem.
  Phys.} \textbf{1984}, \emph{80}, 860\relax
\mciteBstWouldAddEndPuncttrue
\mciteSetBstMidEndSepPunct{\mcitedefaultmidpunct}
{\mcitedefaultendpunct}{\mcitedefaultseppunct}\relax
\EndOfBibitem
\bibitem[Ceriotti \latin{et~al.}(2014)Ceriotti, More, and
  Manolopoulos]{ceri+14cpc}
Ceriotti,~M.; More,~J.; Manolopoulos,~D.~E. {i-PI: A Python interface for ab
  initio path integral molecular dynamics simulations}. \emph{Comp. Phys.
  Comm.} \textbf{2014}, \emph{185}, 1019--1026\relax
\mciteBstWouldAddEndPuncttrue
\mciteSetBstMidEndSepPunct{\mcitedefaultmidpunct}
{\mcitedefaultendpunct}{\mcitedefaultseppunct}\relax
\EndOfBibitem
\bibitem[Ceriotti \latin{et~al.}(2010)Ceriotti, Parrinello, Markland, and
  Manolopoulos]{ceri+10jcp}
Ceriotti,~M.; Parrinello,~M.; Markland,~T.~E.; Manolopoulos,~D.~E. {Efficient
  stochastic thermostatting of path integral molecular dynamics.} \emph{J.
  Chem. Phys.} \textbf{2010}, \emph{133}, 124104\relax
\mciteBstWouldAddEndPuncttrue
\mciteSetBstMidEndSepPunct{\mcitedefaultmidpunct}
{\mcitedefaultendpunct}{\mcitedefaultseppunct}\relax
\EndOfBibitem
\bibitem[Ceriotti and Manolopoulos(2012)Ceriotti, and
  Manolopoulos]{ceri-mano12prl}
Ceriotti,~M.; Manolopoulos,~D.~E. {Efficient First-Principles Calculation of
  the Quantum Kinetic Energy and Momentum Distribution of Nuclei}. \emph{Phys.
  Rev. Lett.} \textbf{2012}, \emph{109}, 100604\relax
\mciteBstWouldAddEndPuncttrue
\mciteSetBstMidEndSepPunct{\mcitedefaultmidpunct}
{\mcitedefaultendpunct}{\mcitedefaultseppunct}\relax
\EndOfBibitem
\bibitem[Senesi \latin{et~al.}(2013)Senesi, Flammini, Kolesnikov, Murray,
  Galli, and Andreani]{sene+13jcp}
Senesi,~R.; Flammini,~D.; Kolesnikov,~A.~I.; Murray,~{\'{E}}.~D.; Galli,~G.;
  Andreani,~C. {The quantum nature of the OH stretching mode in ice and water
  probed by neutron scattering experiments}. \emph{J. Chem. Phys.}
  \textbf{2013}, \emph{139}, 74504\relax
\mciteBstWouldAddEndPuncttrue
\mciteSetBstMidEndSepPunct{\mcitedefaultmidpunct}
{\mcitedefaultendpunct}{\mcitedefaultseppunct}\relax
\EndOfBibitem
\bibitem[Habershon \latin{et~al.}(2009)Habershon, Markland, and
  Manolopoulos]{habe+09jcp}
Habershon,~S.; Markland,~T.~E.; Manolopoulos,~D.~E. {Competing quantum effects
  in the dynamics of a flexible water model.} \emph{J. Chem. Phys.}
  \textbf{2009}, \emph{131}, 24501\relax
\mciteBstWouldAddEndPuncttrue
\mciteSetBstMidEndSepPunct{\mcitedefaultmidpunct}
{\mcitedefaultendpunct}{\mcitedefaultseppunct}\relax
\EndOfBibitem
\bibitem[Kapil \latin{et~al.}(2016)Kapil, Behler, and Ceriotti]{kapi+16jcp2}
Kapil,~V.; Behler,~J.; Ceriotti,~M. {High order path integrals made easy}.
  \emph{J. Chem. Phys.} \textbf{2016}, \emph{145}, 234103\relax
\mciteBstWouldAddEndPuncttrue
\mciteSetBstMidEndSepPunct{\mcitedefaultmidpunct}
{\mcitedefaultendpunct}{\mcitedefaultseppunct}\relax
\EndOfBibitem
\bibitem[Medders \latin{et~al.}(2014)Medders, Babin, and
  Paesani]{medders+jctc2013}
Medders,~G.~R.; Babin,~V.; Paesani,~F. Development of a “First-Principles”
  Water Potential with Flexible Monomers. III. Liquid Phase Properties.
  \emph{Journal of Chemical Theory and Computation} \textbf{2014}, \emph{10},
  2906--2910, PMID: 26588266\relax
\mciteBstWouldAddEndPuncttrue
\mciteSetBstMidEndSepPunct{\mcitedefaultmidpunct}
{\mcitedefaultendpunct}{\mcitedefaultseppunct}\relax
\EndOfBibitem
\bibitem[Behler and Parrinello(2007)Behler, and Parrinello]{behl-parr07prl}
Behler,~J.; Parrinello,~M. {Generalized Neural-Network Representation of
  High-Dimensional Potential-Energy Surfaces}. \emph{Phys. Rev. Lett.}
  \textbf{2007}, \emph{98}, 146401\relax
\mciteBstWouldAddEndPuncttrue
\mciteSetBstMidEndSepPunct{\mcitedefaultmidpunct}
{\mcitedefaultendpunct}{\mcitedefaultseppunct}\relax
\EndOfBibitem
\bibitem[Morawietz \latin{et~al.}(2016)Morawietz, Singraber, Dellago, and
  Behler]{mora+16pnas}
Morawietz,~T.; Singraber,~A.; Dellago,~C.; Behler,~J. {How van der Waals
  interactions determine the unique properties of water}. \emph{Proc. Natl.
  Acad. Sci. USA} \textbf{2016}, \emph{113}, 8368--8373\relax
\mciteBstWouldAddEndPuncttrue
\mciteSetBstMidEndSepPunct{\mcitedefaultmidpunct}
{\mcitedefaultendpunct}{\mcitedefaultseppunct}\relax
\EndOfBibitem
\bibitem[Stephens \latin{et~al.}(1994)Stephens, Devlin, Chabalowski, and
  Frisch]{Stephens1994}
Stephens,~P.~J.; Devlin,~F.~J.; Chabalowski,~C.~F.; Frisch,~M.~J. Ab Initio
  Calculation of Vibrational Absorption and Circular Dichroism Spectra Using
  Density Functional Force Fields. \emph{The Journal of Physical Chemistry}
  \textbf{1994}, \emph{98}, 11623--11627\relax
\mciteBstWouldAddEndPuncttrue
\mciteSetBstMidEndSepPunct{\mcitedefaultmidpunct}
{\mcitedefaultendpunct}{\mcitedefaultseppunct}\relax
\EndOfBibitem
\bibitem[Grimme \latin{et~al.}(2010)Grimme, Antony, Ehrlich, and
  Krieg]{grimme-jcp10-dftd_disp_functional}
Grimme,~S.; Antony,~J.; Ehrlich,~S.; Krieg,~H. A consistent and accurate ab
  initio parametrization of density functional dispersion correction (DFT-D)
  for the 94 elements H-Pu. \emph{The Journal of Chemical Physics}
  \textbf{2010}, \emph{132}, 154104\relax
\mciteBstWouldAddEndPuncttrue
\mciteSetBstMidEndSepPunct{\mcitedefaultmidpunct}
{\mcitedefaultendpunct}{\mcitedefaultseppunct}\relax
\EndOfBibitem
\bibitem[Imbalzano \latin{et~al.}(2018)Imbalzano, Anelli, Giofr\'e, Klees,
  Behler, and Ceriotti]{imba+18jcp}
Imbalzano,~G.; Anelli,~A.; Giofr\'e,~D.; Klees,~S.; Behler,~J.; Ceriotti,~M.
  Automatic selection of atomic fingerprints and reference configurations for
  machine-learning potentials. \emph{The Journal of Chemical Physics}
  \textbf{2018}, \emph{148}, 241730\relax
\mciteBstWouldAddEndPuncttrue
\mciteSetBstMidEndSepPunct{\mcitedefaultmidpunct}
{\mcitedefaultendpunct}{\mcitedefaultseppunct}\relax
\EndOfBibitem
\bibitem[Kapil \latin{et~al.}(2016)Kapil, VandeVondele, and
  Ceriotti]{kapi+16jcp}
Kapil,~V.; VandeVondele,~J.; Ceriotti,~M. {Accurate molecular dynamics and
  nuclear quantum effects at low cost by multiple steps in real and imaginary
  time: Using density functional theory to accelerate wavefunction methods}.
  \emph{J. Chem. Phys.} \textbf{2016}, \emph{144}, 054111\relax
\mciteBstWouldAddEndPuncttrue
\mciteSetBstMidEndSepPunct{\mcitedefaultmidpunct}
{\mcitedefaultendpunct}{\mcitedefaultseppunct}\relax
\EndOfBibitem
\bibitem[Tuckerman \latin{et~al.}(1992)Tuckerman, Berne, and
  Martyna]{tuck+92jcp}
Tuckerman,~M.; Berne,~B.~J.; Martyna,~G.~J. {Reversible multiple time scale
  molecular dynamics}. \emph{J. Chem. Phys.} \textbf{1992}, \emph{97},
  1990\relax
\mciteBstWouldAddEndPuncttrue
\mciteSetBstMidEndSepPunct{\mcitedefaultmidpunct}
{\mcitedefaultendpunct}{\mcitedefaultseppunct}\relax
\EndOfBibitem
\bibitem[Markland and Manolopoulos(2008)Markland, and
  Manolopoulos]{mark-mano08cpl}
Markland,~T.~E.; Manolopoulos,~D.~E. {A refined ring polymer contraction scheme
  for systems with electrostatic interactions}. \emph{Chem. Phys. Lett.}
  \textbf{2008}, \emph{464}, 256\relax
\mciteBstWouldAddEndPuncttrue
\mciteSetBstMidEndSepPunct{\mcitedefaultmidpunct}
{\mcitedefaultendpunct}{\mcitedefaultseppunct}\relax
\EndOfBibitem
\bibitem[Senesi \latin{et~al.}(2013)Senesi, Romanelli, Adams, and
  Andreani]{sene+13cp}
Senesi,~R.; Romanelli,~G.; Adams,~M.~A.; Andreani,~C. {Temperature dependence
  of the zero point kinetic energy in ice and water above room temperature}.
  \emph{Chemical Physics} \textbf{2013}, \emph{427}, 111--116\relax
\mciteBstWouldAddEndPuncttrue
\mciteSetBstMidEndSepPunct{\mcitedefaultmidpunct}
{\mcitedefaultendpunct}{\mcitedefaultseppunct}\relax
\EndOfBibitem
\bibitem[Andreani and Senesi()Andreani, and Senesi]{pcomm01}
Andreani,~C.; Senesi,~R. \relax
\mciteBstWouldAddEndPunctfalse
\mciteSetBstMidEndSepPunct{\mcitedefaultmidpunct}
{}{\mcitedefaultseppunct}\relax
\EndOfBibitem
\bibitem[Cheng \latin{et~al.}(2016)Cheng, Behler, and Ceriotti]{cheng+jcpl2016}
Cheng,~B.; Behler,~J.; Ceriotti,~M. Nuclear Quantum Effects in Water at the
  Triple Point: Using Theory as a Link Between Experiments. \emph{The Journal
  of Physical Chemistry Letters} \textbf{2016}, \emph{7}, 2210--2215, PMID:
  27203358\relax
\mciteBstWouldAddEndPuncttrue
\mciteSetBstMidEndSepPunct{\mcitedefaultmidpunct}
{\mcitedefaultendpunct}{\mcitedefaultseppunct}\relax
\EndOfBibitem
\bibitem[Ceriotti(2014)]{ceri14jpcs}
Ceriotti,~M. {Ab initio simulation of particle momentum distributions in
  high-pressure water}. \emph{Journal of Physics: Conference Series}
  \textbf{2014}, \emph{571}, 12011\relax
\mciteBstWouldAddEndPuncttrue
\mciteSetBstMidEndSepPunct{\mcitedefaultmidpunct}
{\mcitedefaultendpunct}{\mcitedefaultseppunct}\relax
\EndOfBibitem
\bibitem[Medders \latin{et~al.}(2015)Medders, G{\"{o}}tz, Morales, Bajaj, and
  Paesani]{medd+15jcp}
Medders,~G.~R.; G{\"{o}}tz,~A.~W.; Morales,~M.~A.; Bajaj,~P.; Paesani,~F. {On
  the representation of many-body interactions in water}. \emph{J. Chem. Phys.}
  \textbf{2015}, \emph{143}, 104102\relax
\mciteBstWouldAddEndPuncttrue
\mciteSetBstMidEndSepPunct{\mcitedefaultmidpunct}
{\mcitedefaultendpunct}{\mcitedefaultseppunct}\relax
\EndOfBibitem
\bibitem[Reddy \latin{et~al.}(2017)Reddy, Moberg, Straight, and
  Paesani]{reddy2017}
Reddy,~S.~K.; Moberg,~D.~R.; Straight,~S.~C.; Paesani,~F. Temperature-dependent
  vibrational spectra and structure of liquid water from classical and quantum
  simulations with the {MB}-pol potential energy function. \emph{The Journal of
  Chemical Physics} \textbf{2017}, \emph{147}, 244504\relax
\mciteBstWouldAddEndPuncttrue
\mciteSetBstMidEndSepPunct{\mcitedefaultmidpunct}
{\mcitedefaultendpunct}{\mcitedefaultseppunct}\relax
\EndOfBibitem
\bibitem[Paesani(2016)]{Paesani2016}
Paesani,~F. Getting the Right Answers for the Right Reasons: Toward Predictive
  Molecular Simulations of Water with Many-Body Potential Energy Functions.
  \emph{Accounts of Chemical Research} \textbf{2016}, \emph{49},
  1844--1851\relax
\mciteBstWouldAddEndPuncttrue
\mciteSetBstMidEndSepPunct{\mcitedefaultmidpunct}
{\mcitedefaultendpunct}{\mcitedefaultseppunct}\relax
\EndOfBibitem
\end{mcitethebibliography}
\end{document}